\documentclass[prd,12pt,tightenlines,preprintnumbers,showpacs,%
superscriptaddress,nofootinbib,amsmath,amssymb]{revtex4}
\usepackage{graphicx}

\begin{document}

\preprint{CECS-PHY-05-10} \preprint{hep-th/0512074}

\title{Exploring AdS Waves Via Nonminimal Coupling\footnote{Dedicated
to the memory of Professor Jerzy Pleba\'nski}}

\author{Eloy Ay\'on-Beato }\email{ayon-at-cecs.cl}
\affiliation{Centro~de~Estudios~Cient\'{\i}ficos~(CECS),%
~Casilla~1469,~Valdivia,~Chile.}
\affiliation{Departamento~de~F\'{\i}sica,~CINVESTAV--IPN,%
~Apdo.~Postal~14--740,~07000,~M\'exico~D.F.,~M\'exico.}
\author{Mokhtar Hassa\"{\i}ne}\email{hassaine-at-inst.mat.utalca.cl}
\affiliation{Instituto de Matem\'atica y F\'{\i}sica, Universidad de
Talca, Casilla 747, Talca, Chile.}
\affiliation{Centro~de~Estudios~Cient\'{\i}ficos~(CECS),%
~Casilla~1469,~Valdivia,~Chile.}

\begin{abstract}
We consider nonminimally coupled scalar fields to explore the Siklos
spacetimes in three dimensions. Their interpretation as exact
gravitational waves propagating on AdS restrict the source to behave
as a pure radiation field. We show that the related \emph{pure
radiation constraints} single out a unique self-interaction
potential depending on one coupling constant. For a vanishing
coupling constant, this potential reduces to a mass term with a mass
fixed in terms of the nonminimal coupling parameter. This mass
dependence allows the existence of several free cases including
massless and tachyonic sources. There even exists a particular value
of the nonminimal coupling parameter for which the corresponding
mass exactly compensates the contribution generated by the negative
scalar curvature, producing a genuinely massless field in this
curved background. The self-interacting case is studied in detail
for the conformal coupling. The resulting gravitational wave is
formed by the superposition of the free and the self-interaction
contributions, except for a critical value of the coupling constant
where a non-perturbative effect relating the strong and weak regimes
of the source appears. We establish a correspondence between the
scalar source supporting an AdS wave and a \emph{pp} wave by showing
that their respective pure radiation constraints are conformally
related, while their involved backgrounds are not. Finally, we
consider the AdS waves for topologically massive gravity and its
limit to conformal gravity.
\end{abstract}


\maketitle

\section{Introduction}

This year was declared by the UNESCO the international year of
Physics due to the seminal contributions made a century ago by
Einstein to the modern understanding of nature. The legacy of
Einstein not only resides on his successful ideas, he has also been
influential through of its apparent mistakes. His ``biggest
blunder,'' as he called it, was the introduction of a cosmological
constant. However, our view on this subject is different today where
the applications of the ideas related with the cosmological term
rank from quantum field theory to cosmology, and from black holes to
holographic proposals of quantum gravity, just for citing a few
examples.

In this work we shall concentrate in other no-less-important
application: the propagation of gravitational waves in presence of a
cosmological constant. The pioneering studies on this subject
started with the work of Garc\'{\i}a and Pleba\'nski
\cite{Garcia:1981} followed by references
\cite{Salazar:1983,Garcia:1983,Ozsvath:1985qn}. They generalized the
Kundt and Robinson-Trautman vacuum gravitational waves
\cite{Kundt:1961,Robinson:1960} to the case where the cosmological
constant is different from zero (for a review, see
Ref.~\cite{Bicak:1999h}). The spacetimes introduced in
Refs.~\cite{Garcia:1981,Salazar:1983,Garcia:1983,Ozsvath:1985qn} are
algebraically special and contain an interesting symmetrical class
where the multiple principal null direction $k^\mu$, corresponding
to their Weyl tensor, is additionally a Killing vector. This class
is diffeomorphic to the so-called Siklos spacetimes
\cite{Siklos:1985}, which require a negative cosmological constant.

In $D$-dimensions the Siklos spacetimes can be defined by the
following conformal transformation of a \emph{pp} wave background
\begin{equation}
ds^2=\frac{l^2}{y^2}\left[-F(u,y,x^i)du^2-2dudv+dy^2+dx_idx^i\right],
\label{eq:AdSwave}
\end{equation}
where $i$ ranges from $1$ to $D-3$. Here the null Killing field is
given by $k^\mu\partial_\mu=\partial_v$. Note that in four
dimensions, the only Einstein spaces conformally related to a
\emph{pp} wave geometry with a smooth function $F$ are characterized
by the above metric \cite{Siklos:1985}.\footnote{``Impulsive''
\emph{pp} waves, i.e. those allowing a distributional dependence on
retarded time, can be also conformally related to Einstein spaces
with positive curvature \cite{Podolsky:1999sw}.} For a vanishing
structural function $F=0$, we recover the anti-de~Sitter space
metric, while for $F\ll1$ this metric describes just a perturbation
of AdS. In fact, the metric (\ref{eq:AdSwave}) can also be obtained
from the AdS one by a generalized Kerr-Schild transformation
\begin{equation}\label{eq:K-S}
g_{\mu\nu}=g^{\mathrm{AdS}}_{\mu\nu}-\frac{y^2F}{l^2}k_{\mu}k_\nu,
\end{equation}
in an analogous way than \emph{pp} wave backgrounds are obtained
from flat metric by a standard Kerr-Schild transformation. Hence,
the AdS waves are to AdS space what \emph{pp} waves are to Minkowski
space. For a more precise interpretation of the Siklos spacetimes as
exact gravitational waves propagating on AdS space, see
Ref.~\cite{Podolsky:1997ik}.

The wave fronts of these AdS waves, defined by the $(D-2)$-surfaces
$u,v=\mathrm{const.}$, are given by hyperboloids with constant
curvature proportional to $-1/l^2$. Additionally, the null Killing
field $\partial_v$ is geodesic but is not covariantly constant or
parallel. Consequently, the AdS waves are neither \emph{plane}
fronted nor have \emph{parallel} rays as their cousin configurations
the \emph{pp} waves. For this reason we find misleading the term
``AdS \emph{pp} waves'' commonly used in the recent literature to
characterize the gravitational fields (\ref{eq:AdSwave}), and
instead we will refer to them as ``AdS waves'' throughout this
paper. This term is also inaccurate since in general the Siklos
spacetimes are not the only exact gravitational waves propagating on
the AdS background, but we find less harmful to use vague
terminology in comparison to use an incorrect one.

In this paper we are interested in lower dimensional configurations.
For $D=3$, which is the case of our interest, the metric form of the
AdS waves (\ref{eq:AdSwave}) is preserved under the following
coordinate transformations
\begin{equation}\label{eq:formI}
(u,v,y)\mapsto \left(\tilde{u}=\int{\frac{\mathrm{d}u}{f^2}},
\tilde{v}=v-\frac1{2f}\frac{\mathrm{d}f}{\mathrm{d}u}y^2
+\frac12\int\mathrm{d}uF_0, \tilde{y}=\frac{y}f\right),
\end{equation}
together with the redefinition of the structural function
\begin{equation}\label{eq:Fn2F}
\tilde{F}=f^2(F-F_2y^2-F_0).
\end{equation}
Here $f=f(u)$ and $F_0=F_0(u)$ are two arbitrary functions of the
retarded time, and the coefficient $F_2=F_2(u)$ is defined by the
equation
\begin{equation}\label{eq:F2}
F_2=-\frac1{f}\frac{\mathrm{d}^2f}{\mathrm{d}u^2}.
\end{equation}
The above transformations are the $3$-dimensional version of those
found by Siklos in $4$-dimensions \cite{Siklos:1985}. They are
behind of the conformal asymptotic symmetry observed in
asymptotically $AdS_3$ spacetimes \cite{Brown:1986nw}, and can be
generalized to $D$-dimensions and associated to a central extension
of the Virasoro symmetry \cite{Banados:1999tw}.

It follows from the previous transformations that, for a general
function $F$, any quadratic and zeroth order dependence on the
wave-front coordinate $y$ can be locally eliminated by a coordinate
transformation. Indeed, for a given quadratic coefficient $F_2$ we
just need to insert it in the left hand side of Eq.~(\ref{eq:F2})
and to impose that the function $f$ in the coordinate transformation
(\ref{eq:formI}) satisfies the resulting differential equation. We
would like to emphasize that the two spacetimes related by the above
procedure are only equivalent at the local level, since their
corresponding global behavior could be drastically different; in
general the new coordinates in transformation (\ref{eq:formI}) run
over ranges which are different from the ranges of the starting
coordinates. This remark is of exceptional transcendence in $2+1$
dimensions and is one of the lessons exhibited by the discovery of
the BTZ black hole \cite{Banados:wn}, i.e.\ the physically meaning
vacuum configuration in $2+1$ gravity is just a proper
identification of $AdS_3$ \cite{Banados:1992gq}. However, in this
work we just concentrate on local issues and the consequences of the
diverse global behaviors of the AdS waves presented here, in spite
of being an interesting topic by itself, is beyond the scope of the
present paper.

In vacuum $2+1$ AdS gravity, the resulting AdS waves are trivial in
the sense that they can be cast as $AdS_3$ spaces using the above
coordinate transformations. This in turn imply that a matter source
must be introduced in order to support these configurations. In this
work, we show that a nonminimally coupled scalar field presents many
interesting features to be considered as a source. Moreover, this
work is a natural extension of previous ones where the same kind of
matter is nonminimally coupled to \emph{pp} waves but where the
presence of a cosmological constant is generically forbidden
\cite{Deser:2004wd,Ayon-Beato:2004fq,Ayon-Beato:2005bm,Jackiw:2005an}.

The paper is organized as follows. In the next section, we derive
the independent field equations governing the generation of AdS
waves by nonminimally coupled scalar fields. This self-gravitating
process is possible only if the scalar field is constrained by the
fact that all its energy-momentum components vanish except the
energy density along the retarded time, i.e.\ the scalar field must
behave like a pure radiation field. In Sec.~\ref{sec:U}, we show
that the resulting \emph{pure radiations constraints} uniquely
determine the scalar configuration, not only by fixing the
dependence of the field but also by selecting a unique
self-interaction potential for any value of the nonminimal coupling
parameter $\xi$. This potential depends on a single coupling
constant $\lambda$, and for $\lambda=0$ it becomes a mass term. The
related free configurations, studied in Sec.~\ref{sec:free}, are
characterized by the unusual feature that their mass is fixed by the
nonminimal coupling parameter. This section is divided in two parts
where in the first subsection \ref{subsec:m<>0} we determine the AdS
waves corresponding to generic values of the nonminimal coupling
parameter. The second subsection \ref{subsec:m=0} is deserved to the
study of the massless cases, which correspond to the minimal
coupling ($\xi=0$), and the conformal couplings in three ($\xi=1/8$)
and four ($\xi=1/6$) dimensions. Due to the nonminimal coupling to
gravity only the minimal case $\xi=0$ describes a genuinely massless
field on these curved backgrounds while for $\xi=1/8$ and $\xi=1/6$
the mass acquires a tachyonic contribution due to the negative
scalar curvature. For the specific value $\xi=1/5$, this tachyonic
contribution is exactly canceled out and consequently the scalar
field becomes a genuinely massless field on a curved background. In
Sec.~\ref{sec:lamb<>0}, we analyze the case where the scalar field
is self-interacting ($\lambda\not=0$) taking as explicit example the
conformal coupling $\xi=1/8$. In this section we derive the related
AdS wave configuration and show that this later consistently reduces
to the free configuration as the coupling constant $\lambda$ is put
to zero. In fact, the background corresponds to a double Kerr-Schild
transformation of AdS, where the first transformation is just the
free field contribution and the other one depends on the
self-interaction. Additionally, we found a critical value of the
coupling constant given by $\lambda=(\kappa/8l)^2$, for which a
non-perturbative effect relating the strong and weak regimes of the
sources appears. Finally, in Sec.~\ref{sec:AdSw/ppw} we establish a
correspondence between the AdS wave scalar field configurations and
those supporting a \emph{pp} wave. It is shown that, starting from a
\emph{pp} wave scalar source, i.e.\ the scalar field and the
potential which satisfy the pure radiation constraints on the
\emph{pp} wave background, one can generate the AdS wave source and
\emph{vice et versa}. The appendix is devoted to the free case, when
the AdS waves are ruled by topologically massive gravity with a
negative cosmological constant. Conformal gravity configurations are
also obtained as a zero topological mass limit of these
configurations.

\section{\label{sec:source}Nonminimally coupled scalar field
supporting an AdS wave}

In this work we are concerned with scalar fields nonminimally
coupled to an AdS wave background. The field equations are those
arising from the variation of the following action
\begin{equation}
S=\int d^3x\,\sqrt{-g}\left(\frac{1}{2\kappa} (R+2l^{-2})
-\frac{1}{2}\nabla_{\alpha}\Phi\nabla^{\alpha}\Phi-\frac{1}{2}\xi
R\,\Phi^2-U(\Phi)\right), \label{eq:action}
\end{equation}
where $\Lambda=-l^{-2}$ is the negative cosmological constant and
$\xi$ is the parameter allowing a nonminimal coupling to gravity of
the scalar field $\Phi$. The potential $U(\Phi)$ denotes a possible
self-interaction whose form, as we shall see later, is dictated by
the field equations. The variation of the above action with respect
to the metric and the scalar field yield to the Einstein and the
nonlinear Klein-Gordon equations, respectively,
\begin{equation}\label{eq:Ein}
G_{\alpha\beta}-l^{-2}g_{\alpha\beta}={\kappa}T_{\alpha\beta},
\end{equation}
\begin{equation}\label{eq:KG}
\Box\Phi=\xi R\,\Phi+\frac{\mathrm{d}U(\Phi)}{\mathrm{d}\Phi},
\end{equation}
where the energy-momentum tensor is defined by
\begin{equation}\label{eq:emt}
T_{\alpha\beta}=\nabla_{\alpha}\Phi\nabla_{\beta}\Phi
-g_{\alpha\beta}\left(\frac{1}{2}\nabla_{\sigma}\Phi\nabla^{\sigma}\Phi
+U(\Phi)\right)
+\xi\left(g_{\alpha\beta}\Box-\nabla_{\alpha}\nabla_{\beta}
+G_{\alpha\beta}\right)\Phi^2.
\end{equation}

A distinctive feature of an AdS wave lies in the fact that its
Einstein tensor has the following structure
\begin{equation}\label{eq:prad}
G_{\alpha\beta}-l^{-2}g_{\alpha\beta}\propto{k}_{\alpha}k_{\beta},
\end{equation}
which in turn implies that any self-gravitating source supporting
the wave in the presence of the negative cosmological constant must
behave effectively as a pure radiation field \cite{Stephani:2003tm}.
As a direct consequence, in the coordinates of metric
(\ref{eq:AdSwave}), the only component of the Einstein equations
with a nonvanishing left-hand side is the one along $uu$. Thus, the
remaining Einstein equations reduce to the vanishing of the related
energy-momentum tensor components. In what follows, we refer to
these last conditions as the \emph{pure radiation constraints}.

In the present case, we assume that the null Killing field
$k^\mu\partial_\mu=\partial_v$ is also a symmetry of the scalar
field, i.e.\ $\Phi=\Phi(u,y)$. The independent field equations on
the AdS wave background (\ref{eq:AdSwave}) are written using the
following combinations: the equation along the $uu-$component is
given by
\begin{eqnarray}
G_{uu}-l^{-2}g_{uu}-\kappa(T_{uu}-FT_{uv})
             &=&\frac12(1-\kappa\xi\Phi^2)
                y\partial_y\left(\frac1y\partial_yF\right)
\nonumber\\
\label{eq:uu-uv}
             & &{}-\kappa\xi\left(\frac12\partial_y\Phi^2\partial_yF
                -\partial^2_{uu}\Phi^2\right)
                -\kappa(\partial_u\Phi)^2=0,
\end{eqnarray}
while the pure radiation constrains read
\begin{subequations}\label{eq:prc}
\begin{eqnarray}
\label{eq:uy} T_{uy}       &=&\partial_u\Phi\partial_y\Phi
                -\frac{\xi}{y}\partial_y
                \left(y\partial_u\Phi^2\right)=0, \\
\label{eq:uv+yy} T_{yy}+T_{uv}&=&(\partial_y\Phi)^2
                -\frac{\xi}{y^2}\partial_y
                \left(y^2\partial_y\Phi^2\right)=0, \\
\label{eq:yy} T_{yy}
&=&\frac12(\partial_y\Phi)^2-\frac{2\xi}y\partial_y\Phi^2
               -\frac1{y^2}\left[l^2U(\Phi)-\xi\Phi^2\right]=0.
\end{eqnarray}
\end{subequations}
The nonlinear Klein-Gordon equation (\ref{eq:KG}) reduces in this
case to
\begin{equation}\label{eq:KGAdSw}
\frac{y^3}{l^2}\partial_y\left(\frac1y\partial_y\Phi\right)
=-\frac{6\xi}{l^2}\Phi+\frac{\mathrm{d}U(\Phi)}{\mathrm{d}\Phi}.
\end{equation}
Note that since the conservation equations of the energy-momentum
tensor only involve the components given by Eqs.~(\ref{eq:prc}), the
fulfillment of the pure radiation constraints guarantee that the
equation (\ref{eq:KGAdSw}) is automatically satisfied.

>From these equations, it is easy to show the necessity of
introducing a matter source. Indeed in the vacuum case, i.e.
$\Phi=0$, the resulting structural function would be given by
\begin{equation}\label{edq:vacuum}
F(u,y)=F_2(u)y^2-F_0(u),
\end{equation}
and using the coordinate transformation (\ref{eq:formI}), one can
fix $F=0$, and obtaining the metric of AdS.

\section{\label{sec:U}The self-interaction potential}

Here, we explore the possibility of having a self-interacting
nonminimally coupled  ($\xi\ne0$) scalar field acting as a source of
an AdS wave background. As we shall see, the pure radiation
constraints single out the form of the self-interaction potential as
it was also the case in the \emph{pp} wave context
\cite{Ayon-Beato:2005bm}. In order to derive the allowed potential,
it is useful to redefine the scalar field as follows
\footnote{Clearly, the value $\xi=1/4$ deserves a different analysis
which is done at the end of this section.}
\begin{equation}\label{eq:Phi2sigma}
\Phi=\frac1{\sigma^{2\xi/(1-4\xi)}}.
\end{equation}
With this redefinition the pure radiation constraints (\ref{eq:uy})
and (\ref{eq:uv+yy}) are rewritten as
\begin{subequations}\label{eq:s_yyuy}
\begin{eqnarray}
\partial_y\left(y\partial_u\sigma\right)  &=& 0, \\
\partial_y\left(y^2\partial_y\sigma\right)&=& 0,
\end{eqnarray}
\end{subequations}
whose integration yields to
\begin{equation}\label{eq:subssigma}
\sigma(u,y)=\frac{l}{y}\left(\sqrt{\lambda}y+\bar{f}(u)\right),
\end{equation}
where $\lambda$ is a positive constant and $\bar{f}$ is a general
function of the retarded time. Using the expressions
(\ref{eq:Phi2sigma}) and (\ref{eq:subssigma}) it is possible to
rewrite the remaining pure radiation constraint (\ref{eq:yy}) only
in terms of the scalar field. This procedure fixes the
self-interaction potential to be given by
\begin{equation}\label{eq:U(Phi)}
U_\xi(\Phi)=\frac{2\xi\Phi^2}{(1-4\xi)^2}
\left(\xi\lambda\Phi^{(1-4\xi)/\xi}
-\frac{16}{l}\xi(\xi-1/8)\sqrt{\lambda}\Phi^{(1-4\xi)/(2\xi)}
+\frac{24}{l^2}(\xi-1/8)(\xi-1/6)\right).
\end{equation}

Various comments can be made concerning the structure of this
potential. Firstly, for the three-dimensional conformal coupling
$\xi=1/8$, this potential reduces to the conformal one in three
dimensions. Secondly, at the vanishing cosmological constant limit
($l\to\infty$), we recover the potential allowing a self-interacting
scalar field to be nonminimally coupled to a \emph{pp} wave
background \cite{Ayon-Beato:2005bm}.
Finally, it is intriguing that this potential belongs to the same
family of potentials arising in the context of scalar fields
nonminimally coupled to special geometries without inducing
backreaction (the static BTZ black hole \cite{Ayon-Beato:2004ig},
flat space \cite{Ayon-Beato:2005tu}, and the generalized (A)dS
spacetimes \cite{Ayon-Beato:2005c}). Instead, all them has the
common feature that they allow the existence of nontrivial solutions
with a vanishing energy-momentum tensor called ``stealth"
configurations.

We would like to remark that the scalar field configurations given
by Eqs.~(\ref{eq:Phi2sigma}) and (\ref{eq:subssigma}) solve the
nonlinear Klein-Gordon equation (\ref{eq:KGAdSw}) with a
self-interaction potential given by Eq.~(\ref{eq:U(Phi)}) on any AdS
wave background. In other words, this means that the scalar field
and the allowed potential are not sensitive to the structural
function $F$ of the metric (\ref{eq:AdSwave}).

We now derive the self-interaction allowed by the nonminimal
coupling value $\xi=1/4$. In this case we redefine the scalar field
as
\begin{equation}\label{eq:Phi2sigma1/4}
\Phi=\frac1{\sqrt{\kappa}}\mathrm{e}^{\sigma},
\end{equation}
and the pure radiation constraints (\ref{eq:uy}) and
(\ref{eq:uv+yy}) reduce again to the equations (\ref{eq:s_yyuy}).
Hence, we conclude that the solution is given by expression
(\ref{eq:subssigma}), while the remaining pure radiation constraint
(\ref{eq:yy}) gives rise to the following potential
\begin{equation}\label{eq:U1/4(Phi)}
U_{1/4}(\Phi)=\frac{\Phi^2}{4l^2}
\left\{\left[2\ln{\left(\sqrt{\kappa}\Phi\right)}
-l\sqrt{\lambda}+1\right]^2-1\right\}.
\end{equation}

For any value of the nonminimal coupling parameter, we have derived
the allowed potential. In the next section, we explore the existence
of free configurations which are obtained by imposing some
appropriated restrictions on the potential (\ref{eq:U(Phi)}).

\section{\label{sec:free}Free scalar fields}

As shown before, the radiative constraints single out the form of
the potential. For a generic value of the nonminimal coupling
parameter and for a vanishing coupling constant $\lambda=0$, the
potential (\ref{eq:U(Phi)}) becomes a mass term and hence, the
scalar fields can be interpreted as free massive (or massless)
fields. This argument is not valid for $\xi=1/4$ since in this case
the corresponding potential (\ref{eq:U1/4(Phi)}) does not allow the
existence of free fields.

\subsection{\label{subsec:m<>0}Massive cases}

\begin{figure}[h]
\includegraphics[width=8.5cm]{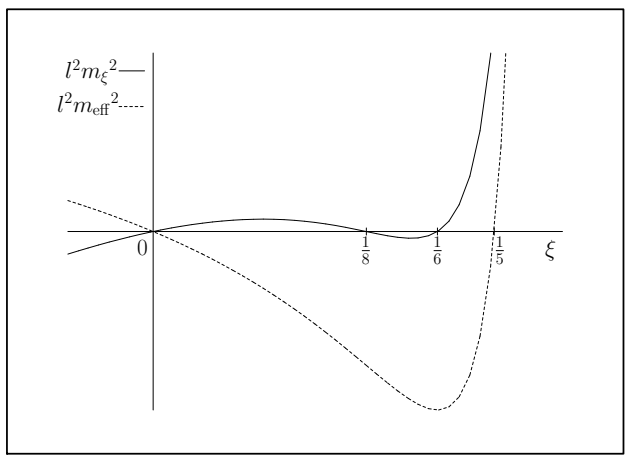}
\caption{\label{fig:mass}The solid graph shows the dependence of the
square of the scalar field mass ${m_\xi}^2$ on the nonminimal
coupling parameter $\xi$, as fixed by the pure radiation
constraints. The dotted graph corresponds to the dependence of the
square of the effective mass ${m_{\mathrm{eff}}}^2$, obtained when
the contribution of the curvature is taken into account, see
Eq.~(\ref{eq:me^2}). The graphs are valid for $\xi\ne1/4$. The
dependence for $\xi>1/4$ is not shown in the graph but it is
positive definite in both cases.}
\end{figure}

For a zero coupling constant $\lambda=0$, the potential
(\ref{eq:U(Phi)}) reduces to a mass term given by
\begin{equation}\label{eq:m^2P^2}
U(\Phi)=\frac12{m_\xi}^2\Phi^2,
\end{equation}
with a mass parameterized in terms of the nonminimal coupling
parameter as
\begin{equation}\label{eq:m^2}
{m_\xi}^2=\frac{6\xi(\xi-1/8)(\xi-1/6)}{l^2(\xi-1/4)^2}.
\end{equation}
It is easy to see that in the minimal case $\xi=0$, and for the
three-dimensional (resp.\ four-dimensional) conformal coupling
parameter, $\xi=1/8$ (resp.\ $\xi=1/6$), this mass vanishes and
their related massless configurations will be analyzed in the next
subsection. On the other hand, the above mass generated by the
nonminimal coupling allows the existence of tachyonic solutions for
negative values of the nonminimal coupling parameter and for
$1/8<\xi<1/6$, as it is shown in FIG.~\ref{fig:mass}.

For the nonminimal couplings associated with a nonvanishing mass the
corresponding massive scalar field satisfying the pure radiation
constraints is given by
\begin{equation}\label{eq:Phim<>0}
\Phi(u,y)=\frac1{\sqrt{\kappa}}
\left(\frac{y}{lf}\right)^{2\xi/(1-4\xi)},
\end{equation}
where we have rescaled appropriately the retarded-time-dependent
function $f$ to be dimensionless. Next, in order to determine the
AdS wave supported by this massive field, we need to solve the
Einstein equation (\ref{eq:uu-uv}). In this case, this equation can
be rewritten as the following hypergeometric differential equation
\begin{equation}\label{eq:hypergeomDE}
x(x-1)\partial^2_{xx}H+\frac{x+2\xi-1}{2\xi}\partial_xH
+\frac{1-4\xi}{2\xi}H=0,
\end{equation}
after making the transformations
\begin{subequations}\label{eq:xH2yF}
\begin{eqnarray}
x&=&\xi\left(\frac{y}{lf}\right)^{4\xi/(1-4\xi)}, \\
H(u,x)&=&
\frac{l^2f^2}{y^2}F+l^2f\frac{\mathrm{d}^2f}{\mathrm{d}u^2}.
\end{eqnarray}
\end{subequations}
The general solution of equation (\ref{eq:hypergeomDE}) reads
\begin{equation}\label{eq:solH}
H(u,x)=F_1(u)\,\,{_2\tilde{F}_1}\!\left(1,\frac{1-4\xi}{2\xi};
                              \frac{1-2\xi}{2\xi};x\right)
      +F_0(u)\left(\frac{x}{\xi}\right)^{(4\xi-1)/(2\xi)},
\end{equation}
where $F_0$ and $F_1$ are two undetermined functions of the retarded
time and ${_2\tilde{F}_1}$ stands for the hypergeometric function
\cite{Erdelyi:1953}. In terms of the original variables it is
possible to apply the coordinate transformation (\ref{eq:formI})
which is equivalent to fix $F_0=0$ and $f=1$. Hence, the AdS waves
supported by massive scalar fields, with a mass (\ref{eq:m^2})
generated by the nonminimal coupling, are given by
\begin{subequations}\label{eq:solm}
\begin{eqnarray}
ds^2&=&\frac{l^2}{y^2}\left[
       -F_1(u)\,\,{_2\tilde{F}_1}\!\left(1,\frac{1-4\xi}{2\xi};
                              \frac{1-2\xi}{2\xi};
       \kappa\xi\Phi^2\right)
       \frac{y^2}{l^2}du^2-2dudv+dy^2\right],\qquad~\\
\label{eq:Phifree} \Phi&=&\frac1{\sqrt{\kappa}}
\left(\frac{y}{l}\right)^{2\xi/(1-4\xi)},
\end{eqnarray}
\end{subequations}
where the function $F_1$ has been properly rescaled.

Note that the above class of AdS wave configurations is not
well-defined for the nonminimal coupling values
\begin{equation}\label{eq:xin}
\xi_n=\frac1{2(1-n)}, \qquad {n}=0,1,2,\ldots,
\end{equation}
since the representation of the hypergeometric function in terms of
the Gauss series is singular for these couplings. The value $n=1$
which corresponds to an infinite nonminimal coupling parameter (or
equivalently to an infinite mass) is excluded in what follows. For
the remaining values $n=0,2,3\ldots,$ the mass is given by
\begin{equation}\label{eq:mn^2}
{m_n}^2=\frac{(n+2)(n+3)}{l^2(n+1)^2(1-n)},
\end{equation}
and the associated solution of Eq.~(\ref{eq:hypergeomDE}) is given
by
\begin{equation}\label{eq:H_n}
H_n(u,x)=\left\{F_1(u)\left[\ln\left(1-\frac{1}{x}\right)
+\sum_{l=1}^{n+1}\frac{1}{lx^l}\right]+F_0(u)\right\}
\left(\frac{x}{\xi_n}\right)^{n+1}.
\end{equation}
Using again the coordinate transformation (\ref{eq:formI}) to
eliminate the functions $F_0$ and $f$, we conclude that for the
nonminimal coupling values $\xi_n=1/[2(1-n)]$ with $n=0,2,3\ldots,$
the solution reads
\begin{subequations}\label{eq:soln}
\begin{eqnarray}
ds^2&=&\frac{l^2}{y^2}\left\{-F_1(u)\left[
\ln\left(1-\frac{1}{\kappa\xi_n\Phi^2}\right)
+\sum_{l=1}^{n+1}\frac{1}{l(\kappa\xi_n\Phi^2)^l}\right]
du^2-2dudv+dy^2\right\},\qquad~\\
\Phi&=&\frac1{\sqrt{\kappa}}\left(\frac{l}{y}\right)^{1/(n+1)}.
\end{eqnarray}
\end{subequations}

\subsection{\label{subsec:m=0}Massless cases}

The simplest massless configuration is the minimal one. In fact for
$\xi=0$, it follows from the pure radiation constraints
(\ref{eq:uv+yy}) and (\ref{eq:yy}), that the theory can not
accommodate a potential. Moreover, the corresponding free massless
scalar field only depends arbitrarily on the retarded time,
$\Phi=\Phi(u)$. In this case, the remaining independent Einstein
equation (\ref{eq:uu-uv}) reduces to the three-dimensional
inhomogeneous Siklos equation \cite{Siklos:1985} for which the
inhomogeneity is dictated by the scalar source,
\begin{equation}
\frac12y\partial_y\left(\frac1y\partial_yF\right)
=\kappa\left(\frac{\mathrm{d}\Phi}{\mathrm{d}u}\right)^2.
\end{equation}
This equation is easily integrated as
\begin{equation}\label{eq:solF0}
F(u,y)=\kappa\left(\frac{\mathrm{d}\Phi}{\mathrm{d}u}\right)^2
\left[\ln{\left(\frac{y}l\right)}+\bar{F_2}(u)\right]y^2 +F_0(u),
\end{equation}
where $F_0$ and $\bar{F_2}$ denote two undetermined functions of the
retarded time. We can now use a special version of the coordinate
change (\ref{eq:formI}) to eliminate the pure quadratic and zeroth
order dependence on the wave-front coordinate $y$. In order to do
that we impose to the function $f$ appearing in the coordinate
change to satisfy the equation (\ref{eq:F2}) with $F_2$ given by
\begin{equation}\label{eq:F22bF2}
F_2=\kappa\left(\frac{\mathrm{d}\Phi}{\mathrm{d}u}\right)^2
\frac{(\bar{F_2}+\ln{f})}{f^4}.
\end{equation}
This choice is motivated by the fact that, after rescaling the
wave-front coordinate, an additional quadratic term appears arising
from the logarithmic function. By doing that, we yield to the
following minimally coupled field content
\begin{subequations}\label{eq:minsol}
\begin{eqnarray}
ds^2&=&\frac{l^2}{y^2}\left[
-\kappa\left(\frac{\mathrm{d}\Phi}{\mathrm{d}u}\right)^2
\ln{\left(\frac{y}l\right)}y^2du^2-2dudv+dy^2\right].\\
\Phi&=&\Phi(u).
\end{eqnarray}
\end{subequations}
Thus, in the minimal case, the scalar field has a wavy behavior
which means that it allows an arbitrary profile in term of the
retarded time, and this profile fully determines the profile of the
AdS wave. Note that this property is also valid in the case of a
\emph{pp} wave configuration with a minimally coupled scalar field
acting as source \cite{Ayon-Beato:2005bm}. The introduction of a
nonminimal coupling induces drastic changes in comparison with the
conclusions of Ref.~\cite{Ayon-Beato:2005bm} regarding \emph{pp}
waves supported by free massless fields. For example, the pure
radiation constraints for the AdS wave geometry are more restrictive
than those for the \emph{pp} wave one since for this later there are
no restrictions on the nonminimal coupling parameter
\cite{Ayon-Beato:2005bm}. Indeed, for the AdS waves the other
massless configurations (apart from the minimal one) are obtained
only for the nonminimal couplings $\xi=1/8$ and $\xi=1/6$, as it can
be concluded from the vanishing of the mass expression
(\ref{eq:m^2}). These two systems are already considered within the
general solution (\ref{eq:solm}), but they are more easily expressed
after fixing the coupling in each case. For the conformal coupling
$\xi=1/8$, the solution is given by
\begin{subequations}\label{eq:solm=01/8}
\begin{eqnarray}
ds^2&=&\frac{l^2}{y^2}\left\{
       -F_1(u)\left[\ln{\left(1-\frac{y}{8l}\right)}+\frac{y}{8l}\right]
       du^2-2dudv+dy^2\right\},\qquad~\\
\Phi&=&\frac1{\sqrt{\kappa}}\sqrt{\frac{y}{l}},
\end{eqnarray}
\end{subequations}
while for the conformal coupling in four dimensions, $\xi=1/6$, the
solution becomes
\begin{subequations}\label{eq:solm=01/6}
\begin{eqnarray}
ds^2&=&\frac{l^2}{y^2}\left[
       -F_1(u)\ln{\left(1-\frac{y^2}{6l^2}\right)}du^2
       -2dudv+dy^2\right],\qquad~\\
\Phi&=&\frac1{\sqrt{\kappa}}\frac{y}{l}.
\end{eqnarray}
\end{subequations}

In contrast with the minimally coupled massless configuration
(\ref{eq:minsol}), the massless sources (\ref{eq:solm=01/8}) and
(\ref{eq:solm=01/6}) have no wavy behavior. The reason is due to the
presence of the cosmological constant which brings through the
curvature an effective mass for any nonminimal coupling
($\xi\not=0$) as it can be seen from the Klein-Gordon equation
(\ref{eq:KGAdSw}). This situation does not occur for the massless
scalar fields nonminimally coupled to \emph{pp} waves
\cite{Ayon-Beato:2005bm} since in this case the scalar curvature is
identically zero.

This last observation motivates the following question: is there
exists some specific values of the nonminimal coupling parameter for
which the generated mass (\ref{eq:m^2}) compensates exactly the
contribution of the cosmological constant to the effective mass? In
this case, this would imply that the resulting configuration is a
truly free massless field. In order to give an answer to this
question, we first define the square of the effective mass as the
sum of the generated mass (\ref{eq:m^2}) together with the
contribution of the scalar curvature,
\begin{equation}\label{eq:me^2}
{m_{\mathrm{eff}}}^2=\xi{R}+{m_\xi}^2
=\frac{5\xi(\xi-1/5)}{4l^2(\xi-1/4)^2}.
\end{equation}
It is clear from this expression that for $\xi=1/5$, which
incidentally corresponds to the six-dimensional conformal coupling,
the effective mass is zero. This in turn means that for $\xi=1/5$,
the pure radiation constraints allow massive fields which behave
effectively as massless scalar fields. The corresponding fields also
belong to the class described by Eq.~(\ref{eq:solm}) and are
expressed as
\begin{subequations}\label{eq:sol1/5}
\begin{eqnarray}
ds^2&=&\frac{l^2}{y^2}\left[
-F_1(u)\mathrm{arctanh}\left(\frac{y^2}{\sqrt{5}l^2}\right)du^2
       -2dudv+dy^2\right],\\
\Phi&=&\frac1{\sqrt{\kappa}}\frac{y^2}{l^2}.
\end{eqnarray}
\end{subequations}
As it was expected, it is easy to see that the above scalar field
satisfies the massless Klein-Gordon equation
\begin{equation}\label{eq:Box}
\Box\Phi=0,
\end{equation}
and this solution is equivalent to the minimal one (\ref{eq:minsol})
in the sense that both describe genuinely massless fields on the
curved AdS-wave background.

As a final remark, we note that configurations that behave
effectively as tachyonic ones exist for values of the nonminimal
parameter between zero and the other effectively-massless coupling
$\xi=1/5$, see FIG.~\ref{fig:mass}.

\section{\label{sec:lamb<>0}Self-interacting scalar fields}

In this section we characterize the AdS waves supported by
self-interacting nonminimally coupled scalar fields. As it has been
shown in Sec.~\ref{sec:U}, the pure radiation constraints fix the
form of the potential which depends on a unique coupling constant
$\lambda$. The properly self-interacting cases correspond to
consider nonvanishing values of this coupling constant. In this
case, the scalar source is described by Eqs.~(\ref{eq:Phi2sigma})
and (\ref{eq:subssigma}) as
\begin{equation}\label{eq:Phi(u,y)}
\Phi(u,y)=\left[\frac{l}y\left(\sqrt{\lambda}\,y+\bar{f}(u)\right)\right]
^{-2\xi/(1-4\xi)}.
\end{equation}
Redefining the structural function by
\begin{equation}\label{eq:xH2yFlamb}
H=F+\frac1{\bar{f}}\frac{\mathrm{d}^2\bar{f}}{\mathrm{d}u^2}y^2,
\end{equation}
the equation (\ref{eq:uu-uv}) which determines the AdS wave
background can be reduced to the following exact form
\begin{equation}\label{eq:Hxxlamb}
\partial_y\left[\frac1y\left(
1-\frac{\kappa\xi}{
\left[\frac{l}y\left(\sqrt{\lambda}y+\bar{f}(u)\right)\right]
^{4\xi/(1-4\xi)}}\right)
\partial_yH\right]=0.
\end{equation}
This equation allows a first integral, and hence the function $F$
can be determined in general in quadratures. The involved analytical
dependence can be expressed in terms of standard functions just for
special values of the nonminimal coupling parameter $\xi$.
Additionally, its general behavior can change for critical values of
the coupling constant $\lambda$. In order to exhibit these features
in a concrete example, we analyze in details the self-interacting
AdS wave configuration for the conformal coupling $\xi=1/8$.

For $\xi=1/8$, the self-interaction potential (\ref{eq:U(Phi)})
reduces to the conformal potential in $2+1$ dimensions
\begin{equation}\label{eq:U1/8(Phi)}
U_{1/8}(\Phi)=\frac\lambda8\Phi^6.
\end{equation}
For a coupling constant $\lambda\not=(\kappa/8l)^2$, it is possible
to redefine the dependence on the retarded time as
$\bar{f}=(\kappa-8l\sqrt{\lambda})f$, where $f$ is a dimensionless
function. In this case, the solution of Eq.~(\ref{eq:Hxxlamb}) turns
out to be
\begin{equation}\label{eq:solH1/8}
H(u,y)=F_1(u)\left[\frac{\sqrt{\lambda}y^2}{16l\kappa{f}^2}
+\frac{y}{8lf}+\ln\left(1-\frac{y}{8lf}\right)\right]+F_0(u).
\end{equation}
Returning to the original structural function by means of
Eq.~(\ref{eq:xH2yFlamb}), and performing the coordinate
transformation (\ref{eq:formI}) it is possible to set $F_0=0$ and
$f=1$. Hence, for a conformal coupling $\xi=1/8$ with a generic
coupling constant $\lambda\ne(\kappa/8l)^2$, the related AdS wave
configuration is given by
\begin{subequations}\label{eq:sol1/8}
\begin{eqnarray}
ds^2&=&\frac{l^2}{y^2}\left\{
       -F_1(u)\left[\frac{\sqrt{\lambda}y^2}{16l\kappa}
       +\frac{y}{8l}+\ln{\left(1-\frac{y}{8l}\right)}\right]
       du^2-2dudv+dy^2\right\},\qquad~\\
\Phi&=&\frac1{\sqrt{\kappa}}
\sqrt{\frac{y}{l}}\left[\frac{8l\sqrt{\lambda}}{\kappa}
\left(\frac{y}{8l}-1\right)+1\right]^{-1/2},
\end{eqnarray}
\end{subequations}
where $F_1$ has been rescaled conveniently. It is interesting to
note that for $\lambda=0$, the free massless configuration
(\ref{eq:solm=01/8}) is straightforwardly obtained. In fact, both
gravitational fields are related by a generalized Kerr-Schild
transformation
\begin{equation}\label{eq:K-Slamb}
g^{\lambda}_{\mu\nu}=g^{\lambda=0}_{\mu\nu}
-\sqrt{\lambda}\frac{F_1y^4}{16l^3\kappa}k_{\mu}k_\nu.
\end{equation}
In other words, this means that the self-interacting configuration
can be described just as a gravitational wave propagating on the
background of the free massless configuration: an exact
gravitational wave on another exact gravitational wave. Since
$g^{\lambda=0}_{\mu\nu}$ is already a Kerr-Schild transformation
from AdS space, expression (\ref{eq:K-Slamb}) coincides with what is
known in the literature as a double Kerr-Schild transformation
\cite{Stephani:2003tm}.

More intriguing is the fact that the manifestation of the
self-interaction arises exactly in a quadratic dependence on the
wave-front coordinate $y$. Until now, it has always be possible and
useful to eliminate locally these contributions by mean of the
coordinate transformation (\ref{eq:formI}). However, in the
self-interacting case there is no isolated quadratic coefficient,
since the profile function $F_1$ is a common coefficient to the
whole wave-front dependence of the structural function $F$ as it can
be seen from Eq.~(\ref{eq:sol1/8}). Thus, it is not judicious to
eliminate this term using the previous arguments, and it seems that
the nonvanishing coupling constant becomes an obstruction to this
local mechanism.

We now analyze the case for which the coupling constant takes the
special value $\lambda=(\kappa/8l)^2$. For a such value, we rewrite
the retarded time dependent function as $\bar{f}=\kappa{f}/8$, where
$f$ is again a dimensionless function. The Eq.~(\ref{eq:Hxxlamb}) is
now solved by
\begin{equation}\label{eq:solH1/8lamb}
H(u,y)=F_1(u)\left(\frac{y^3}{3l^3f^3}+\frac{y^2}{2l^2f^2}\right)+F_0(u).
\end{equation}
By fixing $f=1$ and $F_0=0$ as usual, we end with the following AdS
wave configuration valid for the conformal coupling $\xi=1/8$ and
the special coupling constant $\lambda=(\kappa/8l)^2$,
\begin{subequations}\label{eq:sol1/8lamb}
\begin{eqnarray}
ds^2&=&\frac{l^2}{y^2}\left[
       -F_1(u)\left(\frac{y^3}{3l^3}+\frac{y^2}{2l^2}\right)
       du^2-2dudv+dy^2\right],\qquad~\\
\Phi&=&\sqrt{\frac8{\kappa}\frac{y}{y+l}},
\end{eqnarray}
\end{subequations}
where the function $F_1$ has been rescaled. The above configuration
presents an outstanding non-perturbative feature since the strong
and weak regimes of the source have similar behavior. This can be
shown by first establishing that the fields defined by
\begin{equation}\label{eq:swd}
\hat{\Phi}=\frac8\kappa\frac1\Phi,\quad
\hat{ds}^2=\left(\sqrt{\frac\kappa8}\,\Phi\right)^4ds^2,
\end{equation}
are also solutions of the field equations. Indeed, modulo the
coordinate transformation
\begin{equation}\label{eq:formII}
(u,v,y)\mapsto \left(\hat{u}=u,
\hat{v}=v+\frac1{12}\int\mathrm{d}uF_1,\hat{y}=-y-l\right),
\end{equation}
the new fields can be written as
\begin{subequations}\label{eq:swl}
\begin{eqnarray}
\hat{ds}^2&=&\frac{l^2}{\hat{y}^2}\left[
       F_1(\hat{u})\left(\frac{\hat{y}^3}{3l^3}
       +\frac{\hat{y}^2}{2l^2}\right)
       d\hat{u}^2-2d\hat{u}d\hat{v}+d\hat{y}^2\right],\qquad~\\
\hat{\Phi}&=&\sqrt{\frac8{\kappa}\frac{\hat{y}}{\hat{y}+l}}.
\end{eqnarray}
\end{subequations}
These expressions describe the same solution than the one given by
Eqs.~(\ref{eq:sol1/8lamb}) but with a reflected profile
$\hat{F}_1(\hat{u})=-F_1(\hat{u})$. Since the new and old scalar
fields are inversely proportional, one can conclude a kind of
strong-weak duality where both regimes are diffeomorphic to each
other. Moreover, these two regimes can only be differentiated by the
reflected profiles of their corresponding AdS waves.

Finally, much of the previous analysis can be extended to other
nonminimal couplings. For example for the family of special values
$\xi=m/[4(m+1)]$, $m\in\mathbb{Z}\setminus\{-1\}$, the power
$4\xi/(1-4\xi)$ which appears in Eq.~(\ref{eq:Hxxlamb}) takes the
integer value $m$. In these cases the wave-front dependence is
similar to the dependence obtained in the conformal case
(\ref{eq:solH1/8}). That is, the solution $H$ of the equation
(\ref{eq:xH2yFlamb}) can be written as a common
retarded-time-dependent coefficient multiplying a quadratic term on
the wave-front coordinate, a linear one, and several logarithmic
terms for which the coefficients depend on the roots of a polynomial
of order $m$.

\section{\label{sec:AdSw/ppw} AdS waves \emph{vs} \emph{pp} waves
sources: a Correspondence}

As shown in Sec.~\ref{sec:U}, the AdS wave source, that means the
scalar field $\Phi$ together with its allowed potential $U(\Phi)$,
is completely determined by resolving the pure radiation
constraints. These later which are three of the four independent
Einstein equations do not involve the structural metric function $F$
as it can be verified from their definitions (\ref{eq:prc}). In the
case of a scalar field nonminimally coupled to a \emph{pp} wave, a
similar situation occurs. In the present section, we show that this
is not merely a coincidence and, moreover the analogy can be
extended by establishing a correspondence between the AdS wave and
the \emph{pp} wave sources. In particular, as shown below, starting
from a \emph{pp} wave configuration one can derive the AdS wave
source and \emph{vice et versa}.

In order to make the discussion self-contained we first give a short
review about scalar fields nonminimally coupled to a \emph{pp} wave
\cite{Ayon-Beato:2005bm},
\begin{equation}\label{eq:ppwave}
\bar{g}_{\alpha\beta}dx^{\alpha}dx^{\beta}=-\bar{F}(u,y)du^2-2dudv+dy^2,
\end{equation}
for which the involved field equations are given by
\begin{equation}\label{eq:Einpp}
\bar{G}_{\alpha\beta}=\kappa\bar{T}_{\alpha\beta},
\end{equation}
\begin{equation}\label{eq:KGpp}
\bar{\Box}\bar{\Phi}=\xi \bar{R}\bar{\Phi}
+\frac{\mathrm{d}\bar{U}(\bar{\Phi})}{\mathrm{d}\bar{\Phi}}.
\end{equation}
Since the Einstein tensor of a \emph{pp} wave geometry
(\ref{eq:ppwave}) satisfies
\begin{equation}\label{eq:pradpp}
\bar{G}_{\alpha\beta}\propto{k}_{\alpha}k_{\beta},
\end{equation}
the coupling of any matter source to a \emph{pp} wave also imposes
pure radiation constraints, which in the present case are given by
\begin{subequations}\label{eq:prcpp}
\begin{eqnarray}
\label{eq:uypp} \bar{T}_{uy} &=&
\partial_{u}\bar{\Phi}\partial_{y}\bar{\Phi}
                -\xi\partial_{uy}\bar{\Phi}^2 = 0, \\
\label{eq:uv+yypp} \bar{T}_{yy}+\bar{T}_{uv}
             &=& (\partial_y\bar{\Phi})^2
                -\xi\partial_{yy}\bar{\Phi}^2 = 0,\\
\label{eq:yypp} \bar{T}_{yy} &=& \frac12(\partial_y\bar{\Phi})^2
                -\bar{U}(\bar{\Phi}) = 0.
\end{eqnarray}
\end{subequations}

In Ref.~\cite{Ayon-Beato:2005bm}, we showed that the scalar field is
determined independently of the metric function $\bar{F}$ by solving
the \emph{pp} wave pure radiation constraints (\ref{eq:prcpp}). In
addition, these constraints restrict the self-interaction potential
to be
\begin{equation}\label{eq:U(Phi)pp}
\bar{U}(\bar{\Phi})=\frac{2\xi^2\bar{\lambda}}{(1-4\xi)^2}\bar{\Phi}
                    ^{(1-2\xi)/\xi},
\end{equation}
where $\bar{\lambda}$ is a positive coupling constant.\footnote{The
coupling constant used in Ref.~\cite{Ayon-Beato:2005bm} expressed in
terms of the one used here is $\bar{\lambda}/4$.} To be complete, we
mention that in the free case ($\bar{\lambda}=0$), the scalar field
is shown to be an arbitrary function of the retarded time
$\bar{\Phi}=\bar{\Phi}(u)$, while in the self-interacting case
($\bar{\lambda}\ne0$), it is given by
\begin{equation}\label{eq:Phi(u,y)pp}
\bar{\Phi}(u,y)=\left(\sqrt{\bar{\lambda}}\,y
+\bar{f}(u)\right)^{-2\xi/(1-4\xi)},
\end{equation}
where $\bar{f}$ is an arbitrary function of the retarded time.
Finally, the \emph{pp} wave profile $\bar{F}$ is determined by
solving the remaining independent Einstein equation,
$\bar{G}_{uu}=\kappa \bar{T}_{uu}$.

At this point we would like to stress that the self-interacting
scalar fields supporting the AdS wave (\ref{eq:Phi(u,y)}) and the
\emph{pp} wave (\ref{eq:Phi(u,y)pp}) are functionally related as
\begin{equation}\label{eq:AdS2pp}
\Phi=\Omega^{-2\xi/(1-4\xi)}\bar{\Phi},
\end{equation}
where we have assumed that $\lambda=\bar{\lambda}$. In this
relation, $\Omega=l/y$, and corresponds precisely to the conformal
factor that allows to define the AdS wave metric (\ref{eq:AdSwave})
as a conformal transformation of some \emph{pp} wave background.

Inspired by the previous functional relation, we explore the
possibility of establishing a complete correspondence between the
AdS wave scalar source and the \emph{pp} wave one. More precisely,
we shall prove the following result: starting from the \emph{pp}
wave scalar source, i.e.\ the scalar field (\ref{eq:Phi(u,y)pp}) and
the allowed potential (\ref{eq:U(Phi)pp}), we are able to generate
the AdS wave scalar source described by Eqs.~(\ref{eq:Phi(u,y)}) and
(\ref{eq:U(Phi)}), and \emph{vice et versa}. This is done by
assuming that the scalar fields are conformally related with a
conformal factor given by $\Omega=l/y$ but without fixing the weight
$s$.

In the first part of the next subsection, we show that for a weight
$s=-2\xi/(1-4\xi)$ and for the allowed self-interaction potentials
(\ref{eq:U(Phi)}) and (\ref{eq:U(Phi)pp}) with same coupling
constant, the pure radiation constraints of the AdS wave
(\ref{eq:prc}) are conformally related to those of the \emph{pp}
wave (\ref{eq:prcpp}). Note that this correspondence is strictly
realized between the sources and not between the involved structural
functions, $F$ and $\bar{F}$. However, in the particular situation
of retarded-time-dependent sources, a relation between these
structural functions can be derived. The details are given in the
second part of the next subsection. Finally, the last subsection is
dedicated to configurations that do not depend on the retarded time
for which a more general form of the correspondence can be derived.

\subsection{Off-shell correspondence}

\subsubsection{The sources}

In this subsection, we establish an off-shell correspondence between
the \emph{pp} wave and the AdS wave matter sources. By off-shell, we
mean that the correspondence is realized without using the explicit
form of the scalar field solutions. The expression (\ref{eq:AdS2pp})
suggests to consider a conformal relation between the
self-interacting scalar fields on both gravitational wave
backgrounds as
\begin{equation}\label{eq:sconf}
\Phi=\left(\frac{l}{y}\right)^s\bar{\Phi},
\end{equation}
where the weight $s$ will be fixed later. In this case, the AdS wave
pure radiations constraints (\ref{eq:prc}) can be expressed in terms
of those of the \emph{pp} wave (\ref{eq:prcpp}) as follows
\begin{subequations}\label{eq:prccorr}
\begin{eqnarray}
\label{eq:ert}
T_{uy}&=&\left(\frac{l}{y}\right)^{2s}\left(\bar{T}_{uy}-
\left[s(1-4\xi)+2\xi\right]
\frac{\partial_{u}\bar{\Phi}^2}{2y}\right),\\
\nonumber\\
\label{eq:po} T_{yy}+T_{uv}&=&\left(\frac{l}{y}\right)^{2s}
\left[\bar{T}_{yy}+\bar{T}_{uv} - \left[s(1-4\xi)+2\xi\right]y^{s-1}
\partial_y\left(\frac{\bar{\Phi}^2}{y^s}\right)\right],\quad~\\
\nonumber\\
\label{eq:el}
T_{yy}+\frac{l^2}{y^2}\left[U(\Phi)-V_s(y,\Phi)\right]&=&
\left(\frac{l}{y}\right)^{2s}\left(1+\frac{\sqrt{2}(s+4\xi)\bar{\Phi}}
{y\left[\sqrt{\bar{U}\!\left(\bar{\Phi}\right)}
+\sqrt{\bar{T}_{yy}+\bar{U}\!\left(\bar{\Phi}\right)}\right]}\right)
\bar{T}_{yy},
\end{eqnarray}
\end{subequations}
where the function $V_s(y,\Phi)$ is defined by
\begin{equation}\label{eq:V_s}
V_s(y,\Phi)=
\left(\frac{l}{y}\right)^{\frac{s(4\xi-1)-2\xi}{\xi}}\bar{U}(\Phi)
+\left(\frac{l}{y}\right)^{\frac{s(4\xi-1)-2\xi}{2\xi}}
\frac{\left({s+4\xi}\right)}{l}\Phi\sqrt{2\bar{U}(\Phi)}
+\left(s^2+8\xi s+2\xi\right)\frac{\Phi^2}{2l^2},
\end{equation}
and the functional dependence $\bar{U}$ is given by
Eq.~(\ref{eq:U(Phi)pp}).

>From the first two relations (\ref{eq:ert}) and (\ref{eq:po}), it is
simple to see that the involved pure radiations constraints of both
backgrounds are conformally related if the weight is given by
\begin{eqnarray}\label{eq:weight}
s=-\frac{2\xi}{1-4\xi}.
\end{eqnarray}
In addition, for this value of the weight the function
(\ref{eq:V_s}) looses its dependence on the wave-front coordinate
$y$ and becomes a local function of the scalar field expressed as
\begin{equation}\label{eq:V_sxi}
V_{-2\xi/(1-4\xi)}(y,\Phi)= \bar{U}(\Phi) -
\frac{16\xi\left(\xi-1/8\right)}{(1-4\xi)l}\Phi\sqrt{2\bar{U}(\Phi)}
+\frac{48\xi(\xi-1/8)(\xi-1/6)}{(1-4\xi)^2l^2}{\Phi^2}.
\end{equation}
As it can be seen from the relation (\ref{eq:el}), the remaining
pure radiation constraints are also conformally related provided
that the AdS wave potential $U(\Phi)$ is given by the expression
(\ref{eq:V_sxi}). Note that this expression exactly corresponds to
the potential allowed by the AdS wave source (\ref{eq:U(Phi)}) for
which the coupling constant is taken as $\lambda=\bar{\lambda}$.

In sum, the pure radiation constraints on both gravitational waves
are conformally related with weight $s=-2\xi/(1-4\xi)$, when their
respective potentials (\ref{eq:U(Phi)pp}) and (\ref{eq:U(Phi)}) are
taken with the same coupling constants $\lambda=\bar{\lambda}$,
i.e.\
\begin{subequations}\label{eq:Ts2Tbs}
\begin{eqnarray}
\label{eq:Tuy2Tbuy}
       T_{uy}&=&\left(\frac{l}{y}\right)^{-4\xi/(1-4\xi)}\bar{T}_{uy},\\
\nonumber\\
\label{eq:Tuv2Tbuv}
T_{yy}+T_{uv}&=&\left(\frac{l}{y}\right)^{-4\xi/(1-4\xi)}
                \left(\bar{T}_{yy}+\bar{T}_{uv}\right),\\
\nonumber\\
\label{eq:Tyy2Tbyy}
       T_{yy}&=&\left(\frac{l}{y}\right)^{-4\xi/(1-4\xi)}
\left(1-\frac{16\sqrt{2}\xi(\xi-1/8)\bar{\Phi}}
{(1-4\xi)y\left[\sqrt{\bar{U}\!\left(\bar{\Phi}\right)}
+\sqrt{\bar{T}_{yy}+\bar{U}\!\left(\bar{\Phi}\right)}\right]}\right)
\bar{T}_{yy}.\quad~
\end{eqnarray}
\end{subequations}
Using these relations, it is possible to generate any AdS wave
scalar source from the \emph{pp} wave source through a conformal
transformation with a particular weight depending on the nonminimal
coupling parameter, and considering the same coupling constant in
both cases $\lambda=\bar{\lambda}$. For example, for
$\bar{\lambda}\ne0$, the self-interacting scalar solutions
supporting the \emph{pp} waves also generate the self-interacting
sources ($\lambda\ne0$) which support the AdS waves. On the other
hand, for $\bar{\lambda}=0$ the free massless \emph{pp} wave scalar
fields depend only on the retarded time and allow to generate the
free massive and massless configurations ($\lambda=0$) supporting
the AdS waves analyzed in Sec~\ref{sec:free}.

Finally, we note that similar conclusions can be obtained by
considering the scalar wave equations. Using the conformal relation
(\ref{eq:sconf}) with the weight (\ref{eq:weight}) and assuming
$\lambda=\bar{\lambda}$, the Klein-Gordon equations in both
backgrounds are not conformally related in general. However, a
conformal relation can be reached for the following combination
involving the wave equations
\begin{eqnarray}
2\xi\Phi\left(\Box\Phi-\xi
R\Phi-\frac{\mathrm{d}U(\Phi)}{\mathrm{d}\Phi}\right)-g^{yy}T_{yy}&=&
\left(\frac{l}{y}\right)^{2(2\xi-1)/(1-4\xi)}\nonumber\\
&&{}\times\left[2\xi\bar{\Phi}\left(\bar{\Box}\bar{\Phi}
-\xi\bar{R}\bar{\Phi}
-\frac{\mathrm{d}\bar{U}(\bar{\Phi})}{\mathrm{d}\bar{\Phi}}\right)
-\bar{g}^{yy}\bar{T}_{yy}\right].\quad~ \label{eq:weads'''}
\end{eqnarray}
It is clear from this expression that any solution of the wave
equation in one background satisfying also the pure radiation
constraints must be necessarily a solution of the wave equation in
the other background. A particularity obviously occurs for the
conformal coupling $\xi=1/8$, since in this case one can use the
relation (\ref{eq:Tyy2Tbyy}) to eliminate the energy-momentum
components from the above combinations, yielding to the standard
conformal correspondence between the wave equations in this case. We
never have seen a relation like the above in the literature, it has
sense only for strictly nonminimal couplings ($\xi\ne0$) and
gravitational waves backgrounds, but it would be very interesting if
there exist other geometries allowing its existence.

\subsubsection{The backgrounds}

In the above treatment we have established a correspondence between
the scalar sources supporting a \emph{pp} wave and an AdS wave. Here
we shall show that in the case of scalar fields depending on the
retarded time it is also possible to build a relation between both
gravitational waves. Up to now in this section, the only unexplored
component of the Einstein equations has been the $uu$ one, since it
does not participate in the pure radiations constraints. We shall
use it now, due to it is the only component containing information
on the metric structural functions.

Using the conformal relation (\ref{eq:sconf}) and eliminating the
second derivatives of the scalar field with respect to the retarded
time from the $uu$ components of the Einstein equations in both
backgrounds the following relation is obtained
\begin{equation}\label{eq:backg_uu}
E_{uu}-FE_{uv} - \frac{y}2\partial_y\left(
\left(1-\kappa\xi\Phi^2\right) \frac{\partial_yF}{y}\right)
=\left(\frac{l}{y}\right)^{2s}\left(\bar{E}_{uu}-\bar{F}\bar{E}_{uv}
-\frac12\partial_y\left[\left(1-\kappa\xi\bar{\Phi}^2\right)
\partial_y\bar{F}\right]\right),
\end{equation}
where $E_{\alpha\beta}$ and $\bar{E}_{\alpha\beta}$ denote the
components of the Einstein equations for the AdS and \emph{pp} wave
cases, respectively, i.e.\ they are defined by
\begin{eqnarray}\label{eq:Es}
E_{\alpha\beta}      &=&G_{\alpha\beta}-l^{-2}g_{\alpha\beta}
                        -{\kappa}T_{\alpha\beta},\\
\bar{E}_{\alpha\beta}&=&\bar{G}_{\alpha\beta}
                        -\kappa\bar{T}_{\alpha\beta}.
\end{eqnarray}
As consequence of the identity (\ref{eq:backg_uu}) a \emph{pp} wave
solution ($\bar{E}_{\alpha\beta}=0$) implies the existence of an AdS
wave solution ($E_{\alpha\beta}=0$), and viceversa, iff both
structural functions satisfy a differential equation which allows to
determine one solution in terms of the other. For example, for
obtaining the AdS wave solution the differential equation is given
by
\begin{equation}\label{eq:deq_F2bF}
y\partial_y\left\{
\left[1-\kappa\xi\left(\frac{y}{l}\right)^{4\xi/(1-4\xi)}\bar{\Phi}^2
\right]
\frac{\partial_yF}{y}\right\}=\left(\frac{y}{l}\right)^{4\xi/(1-4\xi)}
\partial_y\left[\left(1-\kappa\xi\bar{\Phi}^2\right)
\partial_y\bar{F}\right],
\end{equation}
where we are additionally using that the remaining Einstein
equations imply that $s=-2\xi/(1-4\xi)$ as was shown previously.
Finally, the above equation allows to determine in quadratures the
AdS wave structural function $F$ in terms of the \emph{pp} wave
solution
\begin{eqnarray}
F(u,y)&=&\int{\frac{\mathrm{d}y\,y}
{\left(1-\kappa\xi\left({y}/{l}\right)^{4\xi/(1-4\xi)}
\bar{\Phi}^2\right)} \biggl[F_1(u)+\int{\frac{\mathrm{d}y}l
\left(\frac{y}{l}\right)^{(8\xi-1)/(1-4\xi)}
\partial_y\left[\left(1-\kappa\xi\bar{\Phi}^2\right)
\partial_y\bar{F}\right]}\biggr]}\nonumber\\
&&{} +F_0(u). \label{eq:F2bF}
\end{eqnarray}

In the case of scalar configurations independent of the retarded
time the relation (\ref{eq:backg_uu}) can not be achieved. It
appears that there is no relation between the backgrounds in this
case.

\subsection{On-shell correspondence}

In the AdS wave case and for a generic nonminimal coupling parameter
($\xi\not=0$), the retarded-time-dependent integration functions of
the scalar field can always be put equal to some non zero constant
by an appropriate coordinate transformation. In the case of the
\emph{pp} wave configurations, the situation is different. Indeed,
as shown in Ref.~\cite{Ayon-Beato:2005bm}, the arbitrary retarded
time dependent functions can only be removed for a self-interacting
scalar field. In this case, a suitable shift in the wave-front
coordinate $y$ allows to remove the integration function $\bar{f}$
from (\ref{eq:Phi(u,y)pp}) leading to the following ``physical''
solution
\begin{equation}\label{eq:phymi}
\bar{\Phi}=\left(\sqrt{\bar{\lambda}}\,y\right)^{-2\xi/(1-4\xi)}.
\end{equation}

In view of these remarks, it is not clear that the correspondence
reported previously is still valid once the undetermined integration
functions have been removed. In fact, we shall see that the two
systems, namely the self-interacting scalar fields nonminimally
coupled to a \emph{pp} wave and the nonminimal configurations
supporting the AdS wave (once the integration functions have been
removed in each case) can also be linked through a more general
correspondence.

We start by looking for a more general relation between the scalar
fields,
\begin{eqnarray}
\Phi=H(\Omega)^{-2\xi/(1-4\xi)}\,\bar{\Phi}, \label{gconf}
\end{eqnarray}
where $H$ is a function of the conformal factor $\Omega=l/y$ and
$\bar{\Phi}$ is the physical \emph{pp} wave scalar field
(\ref{eq:phymi}). In this case, the AdS wave pure radiations
constraints (\ref{eq:prc}) reduce to
\begin{subequations}\label{eq:prccorr2}
\begin{eqnarray}
\label{eq:ert'}
       T_{uy}&=&0,\\
\nonumber\\
\label{eq:po'}
T_{uv}+T_{yy}&=&\frac{4\xi^2\Omega^5\Phi^2}{l^2(1-4\xi)H}
\,\frac{\mathrm{d}^2}{\mathrm{d}\Omega^2}\!\left(\frac{H}\Omega\right),\\
\nonumber\\
\nonumber
T_{yy}&=&-\Omega^2U(\Phi)+\frac{\xi\Phi^2\Omega^2}{H^2(4\xi-1)^2\,l^2}
\\
&&{}\times
\left[2\xi\Omega^2\left(\frac{\mathrm{d}H}{\mathrm{d}\Omega}\right)^2
+4\xi(8\xi-3)\Omega H\frac{\mathrm{d}H}{\mathrm{d}\Omega}
+(1+2\xi-16\xi^2)H^2\right],\label{eq:el'}
\end{eqnarray}
\end{subequations}
where in order to derive these relations, we have make use of the
explicit form of the scalar field solution nonminimally coupled to a
\emph{pp} wave, (\ref{eq:phymi}).

The vanishing of the second equation (\ref{eq:po'}) implies that the
function $H$ is given by
\begin{equation}\label{eq:H}
H(\Omega)=\alpha \Omega+\beta \Omega^2,
\end{equation}
where $\alpha$ and $\beta$ are two arbitrary constants. Combining
this expression with the explicit form of the scalar field solution
(\ref{eq:phymi}) on a \emph{pp} wave background, we find that the
AdS wave scalar field can be written as
\begin{equation}\label{eq:Phi2ab}
\Phi=\left[\frac{l}y\left(\alpha\sqrt{\bar{\lambda}}\,y
+\beta\sqrt{\bar{\lambda}}l\right)\right]^{-2\xi/(1-4\xi)}
=\left[\sqrt{\bar{\lambda}}l
\left(\alpha+\beta\Omega\right)\right]^{-2\xi/(1-4\xi)}.
\end{equation}
This dependence is identical to the solution given in
Eq.~(\ref{eq:Phi(u,y)}) where $\alpha^2\bar{\lambda}$ plays now the
role of the coupling constant and the function depending on the
retarded time is given by the constant $\beta\sqrt{\bar{\lambda}}l$.
In fact, using the above expression in the vanishing of
Eq.~(\ref{eq:el'}) one recovers the self-interacting potential
(\ref{eq:U(Phi)}) allowing the AdS wave configurations, but with a
coupling constant given now by $\lambda=\alpha^2\bar{\lambda}$. This
reflects the fact that by construction, the equations determining
the AdS wave source, namely the pure radiation constraints and the
nonlinear Klein-Gordon equation, are invariant under the rescaling
$\lambda\to \alpha^2\lambda$. This is a consequence of the fact that
the scalar contribution to action (\ref{eq:action}) only change by a
multiplicative constant under the above rescaling. As it was
previously anticipated, the free constant $\beta$ corresponds to the
nonzero constant remaining in the scalar field once the
retarded-time-dependent integration function is removed. It can be
fixed appropriately using the transformations (\ref{eq:formI}), as
it has been explicitly shown in Sec.~\ref{sec:lamb<>0}.

For $\alpha\ne0$ the present correspondence is analog to the
correspondence established in the previous subsection, in the sense
that it allows to relate the self-interacting configurations on a
\emph{pp} wave (i.e.\ $\bar{\lambda}\ne0$) with the self-interacting
configurations on an AdS wave (i.e.\
$\lambda=\alpha^2\bar{\lambda}\ne0$). The basic difference lies in
the fact that the integration functions depending on retarded time
are appropriately fixed in each case. However, for $\alpha=0$, a new
link can be made. Indeed, in this case, the coupling constant in the
AdS wave side vanishes ($\lambda=\alpha^2\bar{\lambda}=0$) and as a
consequence, the potential reduces to the mass term given by
Eqs.~(\ref{eq:m^2P^2}) and (\ref{eq:m^2}). Hence, the free scalar
fields on an AdS wave (\ref{eq:Phifree}) and the self-interacting
scalar fields on a \emph{pp} wave (\ref{eq:phymi}) are effectively
related provided the constant $\beta$ to be fixed by
$$
\beta=\left(\sqrt{\bar{\lambda}}\,l\right)^{-1}
      \kappa^{(1-4\xi)/(4\xi)}.
$$
In summary, we have generated the free massive and massless
configurations ($\lambda=0$) supporting an AdS wave from the
self-interacting \emph{pp} wave configurations
($\bar{\lambda}\ne0$).

\section{\label{sec:conclu}Conclusions}

In this paper, we have attacked the problem of the generation of
exact gravitational waves propagating on AdS space. We have
restricted our attention to the three-dimensional case for which a
matter source must be necessarily introduced in order to support
these AdS waves. The elaboration of this work has opened many
interesting questions that go beyond the simple mathematical
resolution.

The first interrogation may concern the choice of the matter source
for these AdS waves. Indeed, due to the wave character of these
gravitational fields, any matter supporting them must behave as a
pure radiation field. This means that the only nonvanishing
component of the energy momentum tensor must be the energy density
along the retarded time. This last fact is certainly very
restrictive concerning the possible choices of matter source. It is
interesting to remark that a source given by a scalar field
nonminimally coupled to the AdS waves, as we have considered here,
does not yield to inconsistencies, quite the contrary provides many
interesting curiosities that we have reported throughout this paper.
Although it is also natural to ask about the motivation of
considering such source instead of another plausible one, the
appearance of these unexpected curiosities (essentially due to the
inclusion of the nonminimal coupling) in some sense legitimate our
choice. As the first curiosity, it is intriguing that the
integration of the \emph{pure radiation constraints}, which are
three of the four independent Einstein equations, completely fixes
the dependence of the scalar field as well as singles out a unique
self-interaction potential allowing the existence of AdS waves. In
other words, this means that the self-interacting nature of the
source is strictly determined from the requirement of being capable
of generating AdS waves.

We showed that the resulting potential depends only on one coupling
constant denoted by $\lambda$ and presents various interesting
characteristics. Its expression is given by a superposition of
different powers of the scalar field whose exponents are expressed
in terms of the nonminimal coupling parameter $\xi$. For the
three-dimensional conformal coupling, i.e.\ $\xi=1/8$, the
parameterized expression of the potential reduces precisely to the
three-dimensional conformal potential $U_{1/8}(\Phi)\propto\Phi^6$.
This in turn implies that the matter source allowed by the AdS waves
becomes conformally invariant. It is amusing to note that there is
\emph{\`a priori} no good reason for the potential to become the
conformally invariant one, since the full system does not exhibit
the conformal invariance. We would like to remark that this is not
the first time that the above nonminimal-coupling related potential
appears. In the limit of vanishing cosmological constant
$l\to\infty$ the potential becomes the one allowing the existence of
scalar-field generated \emph{pp} waves derived in
Ref.~\cite{Ayon-Beato:2005bm}. Additionally, it is a particular case
of the family of potentials allowing the existence of
gravitationally stealth configurations on special backgrounds as the
static BTZ black hole, flat space, and (A)dS, see
Refs.~\cite{Ayon-Beato:2004ig,Ayon-Beato:2005tu,Ayon-Beato:2005c}.

After determining the scalar source from the pure radiation
constraints, the integration of the remaining Einstein equation
permits to find the geometric background. The simplest case occurs
for a vanishing coupling constant $\lambda=0$, since the potential
reduces to a pure mass term with the peculiarity that the
corresponding mass ${m_\xi}^2$ is fixed in terms of the nonminimal
coupling parameter with a mass scale determined from the
cosmological constant. This last fact increases the interest on the
nonminimal coupling parameter $\xi$ whose range now allows the
existence of several types of free sources. For example, for
$1/8<\xi<1/6$ or $\xi<0$, the AdS waves are generated by tachyonic
configurations (${m_\xi}^2<0$). The waves can also be supported by
massless fields in the cases of the minimal coupling $\xi=0$, the
three-dimensional conformal coupling $\xi=1/8$, and the
four-dimensional conformal coupling $\xi=1/6$. However, the last two
cases do not represent genuinely massless fields in their
corresponding curved backgrounds. This is due to the fact that there
is an additional contribution to their mass coming from de
nonminimal coupling of the fields to the gravitational waves. Since
the AdS waves has negative scalar curvature this contribution is
tachyonic, as a consequence the fields acquire an effective mass
given by ${m_{\mathrm{eff}}}^2=-\xi6l^{-2}+{m_\xi}^2$ on these
backgrounds. The explicit dependence of the effective mass on the
nonminimal coupling parameter implies that in the range $0<\xi<1/5$
the scalar fields are forced to behave effectively as tachyonic
fields (${m_{\mathrm{eff}}}^2<0$). As it can be anticipated from
continuity, the genuinely massless states (${m_{\mathrm{eff}}}^2=0$)
are not only realized for the minimal coupling $\xi=0$ but also for
the nonminimal coupling value $\xi=1/5$, which incidentally
corresponds to the conformal coupling in $D=6$. The value $\xi=1/5$
is associated to a critical case in the sense that its allowed mass
${m_{1/5}}^2$ exactly compensates the contribution generated by the
negative cosmological constant, this explain why the scalar field
becomes a truly massless field in a curved background. It is true
that the matter source is characterized by a unique parameter,
namely the nonminimal coupling parameter, but once again there is no
good reason for the existence of a critical value of this parameter
that renders the scalar field effectively as a massless field.
Additionally, the fact that the fine-tuning occurs for the conformal
coupling of a higher dimension seems to indicate that the effect can
be connected to some more symmetrical higher-dimensional physics. An
interesting work will consist in exploring if this mass annihilating
effect can be extended to higher-dimensional versions of the AdS
waves or to other geometries.

The self-interacting case, i.e.\ $\lambda\ne0$, has been studied in
details for the conformal coupling $\xi=1/8$. Generically, the
background have been shown to be a gravitational wave propagating on
AdS space builds from the superposition of two contributions. The
first one corresponds to the massless free field ($\lambda=0$). The
other contribution is associated with the self-interaction, since it
presents a dependence on the coupling constant which goes as
$\sim\sqrt{\lambda}$. As a consequence, the free configuration is
consistently obtained from the self-interacting one in the limit
$\lambda\to0$. This double Kerr-Schild representation of the
background starting from AdS space is outstanding, since in spite of
the strong self-coupling of the field ($\propto\Phi^6$) its
contribution as source is encoded in a very simple way. This simple
connection between the free and the self-interacting cases occurs
only if the coupling constant is different from the critical value
$\lambda=(\kappa/8l)^2$. For this critical value, we have proved
that a surprising non-perturbative effect occurs. We have
established a kind of strong-weak duality in the sense that the
field strengths $\Phi$ and $1/\Phi$ are diffeomorphic and the
profile of their corresponding AdS waves only differs by a minus
sign. In other words the strong regime is locally equivalent to the
weak one just by reflecting the profile of the gravitational wave.
These results for the conformal coupling can be generalized to the
family of nonminimal coupling values $\xi=m/[4(m+1)]$,
$m\in\mathbb{Z}\setminus\{-1\}$, and in the same line as in the
conformal case, different critical behaviors can also appear.

Another intriguing fact concerns the relation that we have
established with the problem of scalar fields nonminimally coupled
to a \emph{pp} wave. It is well-known that an AdS wave can also be
viewed as a conformal transformation of a \emph{pp} wave but there
is no reason for their matter sources to be also in correspondence.
Here, we have shown that the pure radiation constraints of both
systems are conformally related when their scalar fields are also
conformally related with a conformal weight fixed in terms of the
nonminimal coupling parameter. This conformal correspondence occurs
between the scalar sources (scalar fields and allowed potentials in
each case) but the correspondence can not be extended to the
involved backgrounds. In addition, we have shown that a particular
combination of the Klein-Gordon equation together with the
wave-front component of the energy-momentum tensor $T_{yy}$ is in
fact conformally invariant in the pass from one gravitational wave
to the other. In the case of the conformal coupling this relation
becomes the well-known conformal invariance of the conformal
Klein-Gordon equation. It will be interesting to explore the
transcendence of similar conclusions in higher dimensions.

Finally, we have considered the case for which the AdS waves are
governed by topologically massive gravity with a negative
cosmological constant. Unlike standard $2+1$ gravity, the waves are
allowed in the vacuum case. The negative cosmological constant acts
on the gravitational waves by reducing the value of their
topological mass, i.e.\ if the theory has a topological mass $\mu$
the AdS waves have a physical topological mass
$\mu_{\mathrm{eff}}=\mu\sqrt{1-(l\mu)^{-2}}$. These vacuum
configurations coincide with the ones derived in
Ref.~\cite{Ayon-Beato:2004fq} from a correspondence with conformal
gravity. For the critical values of the topological mass
$\mu=\pm{l}^{-1}$, the corresponding AdS waves can not be
interpreted as massive Klein-Gordon modes, which seems to indicate
that in general the theory has no massive character in these limits.
The above considerations have been extended to waves supported by
free nonminimally coupled scalar fields. In the minimal case the
gravitational wave is just a superposition of two contributions
given by the vacuum AdS wave and by the AdS wave corresponding to
Einstein gravity supported by a minimally coupled scalar field with
an effective gravitational constant
$\kappa_{\mathrm{eff}}=\kappa/[1+(l\mu)^{-1}]$. We have also studied
the zero topological mass limit of the above nonvacuum
configurations and showed that they become the AdS wave solutions of
conformal gravity.

\begin{acknowledgments}
We thank D.~Correa, C.~Mart\'{\i}nez, M.~Ortaggio, R.~Troncoso, and
J.~Zanelli for useful discussions and especially A. Garcia. This
work is partially supported by grants 3020032, 1040921, 7040190,
1051064, 1051084, and 1060831 from FONDECYT, grants 38495E, 34222E,
and CO1-41639 from CONACyT, and grant 2001-5-02-159 from
CONICYT/CONACyT. Institutional support to the Centro de Estudios
Cient\'{\i}ficos (CECS) from Empresas CMPC is gratefully
acknowledged. CECS is a Millennium Science Institute and is funded
in part by grants from Fundaci\'{o}n Andes and the Tinker
Foundation.
\end{acknowledgments}

\appendix*

\section{\label{app:TMG}AdS waves for Topologically Massive Gravity
with a Cosmological Constant}

In this Appendix, we analyze the AdS waves ruled by topologically
massive gravity \cite{Deser:1981wh}, when this theory is
supplemented with a negative cosmological constant. The scalar
configurations nonminimally coupled to the \emph{pp} waves of this
theory were studied in
Refs.~\cite{Deser:2004wd,Ayon-Beato:2004fq,Ayon-Beato:2005bm,Jackiw:2005an}
using different perspectives. The topologically massive gravity
modifies the Einstein equations (\ref{eq:Ein}) by the addition of
the Cotton tensor. However, for the AdS wave metric
(\ref{eq:AdSwave}) the corresponding left hand side is again
constrained to satisfy
\begin{equation}\label{eq:pradTMG}
\frac1{\mu}C_{\alpha\beta}+G_{\alpha\beta}-l^{-2}g_{\alpha\beta}
\propto{k}_{\alpha}k_{\beta},
\end{equation}
where $C_{\alpha\beta}$ is the Cotton tensor and $\mu$ is the
so-called topological mass, see
Refs.~\cite{Deser:2004wd,Ayon-Beato:2004fq,Ayon-Beato:2005bm,Jackiw:2005an}
for definitions and conventions. The above relation is due to the
fact that the only nonvanishing component of the Cotton tensor is
given by $C_{uu}=y/(2l)\partial^3_{yyy}F$, and consequently the
field equations involve the same pure radiation constraints
(\ref{eq:prc}). In what follows, we analyze different aspects of
scalar configurations associated with the topologically massive
gravity.

\subsection{Vacuum AdS waves}

Let us first consider the vacuum case $T_{\alpha\beta}=0$ for which
the only nontrivial equation is the $uu$ one
\begin{equation}\label{eq:TMGveq}
\frac1{\mu}C_{uu}+G_{uu}-l^{-2}g_{uu}
=\frac{y}{2l\mu}\partial_y\left(
\frac1{y^{l\mu}}\partial_y\left(y^{l\mu}\partial_yF\right)\right)=0.
\end{equation}
The solution in this case, after an appropriate coordinate change in
the same line of those used in the whole paper, reads
\begin{equation}\label{eq:TMGAdSwv}
ds^2=\frac{l^2}{y^2}\left[-F_1(u)\left(\frac{y}l\right)^{1-l\mu}du^2
-2dudv+dy^2\right].
\end{equation}
The vacuum equation (\ref{eq:TMGveq}) allows two other solutions for
the special values of the topological mass $\mu=\pm{l}^{-1}$, which
are given in each case as
\begin{equation}\label{eq:TMGvmul1}
ds^2=\frac{l^2}{y^2}\left[
-F_1(u)\ln{\left(\frac{y}l\right)}du^2-2dudv+dy^2\right],
\end{equation}
for $\mu={l}^{-1}$, and
\begin{equation}\label{eq:TMGvmul-1}
ds^2=\frac{l^2}{y^2}\left[
-F_1(u)\ln{\left(\frac{y}l\right)}y^2du^2-2dudv+dy^2\right],
\end{equation}
for $\mu=-{l}^{-1}$. The AdS waves (\ref{eq:TMGAdSwv}) and
(\ref{eq:TMGvmul-1}) were derived previously in
Ref.~\cite{Ayon-Beato:2004fq} using a correspondence established in
that work between the conformal gravity with a self-interacting
conformal scalar source on one side and the topologically massive
gravity with a negative cosmological constant on the other side.
Their corresponding configurations in conformal gravity are
\emph{pp} wave backgrounds. The solution (\ref{eq:TMGvmul1}) has no
analog in the conformal gravity side, since it would correspond to a
negative gravitational constant in this theory, see
Ref.~\cite{Ayon-Beato:2004fq}.

The structural function of the generic vacuum AdS waves
(\ref{eq:TMGAdSwv}) satisfies the Klein-Gordon equation
\begin{equation}\label{eq:K_Gmu}
\Box{F}={\mu_{\mathrm{eff}}}^2F,
\end{equation}
with an effective mass defined by
$\mu_{\mathrm{eff}}=\mu\sqrt{1-(l\mu)^{-2}}$. This means that the
effect of the negative cosmological constant is to lower the value
of the physical topological mass. For the special values
$\mu=\pm{l}^{-1}$ the corresponding structural functions in the AdS
waves (\ref{eq:TMGvmul1}) and (\ref{eq:TMGvmul-1}) do not satisfy a
Klein-Gordon equation, which indicates that the theory does not have
a massive character for these critical values.

\subsection{AdS waves for a minimal scalar field}

We study now the AdS waves of topologically massive gravity for
which the source is a minimally coupled scalar field. For $\xi=0$
the pure radiation constraints (\ref{eq:prc}) imply that the scalar
field only depends on the retarded time, $\Phi=\Phi(u)$, and
additionally there is no self-interaction in this case, $U(\Phi)=0$.
The remaining Einstein equation is given by the $uu-$component
\begin{equation}\label{eq:TMGuuxi0}
\frac{y}{2l\mu}\partial_y\left(
\frac1{y^{l\mu}}\partial_y\left(y^{l\mu}\partial_yF\right)\right)
=\kappa\left(\frac{\mathrm{d}\Phi}{\mathrm{d}u}\right)^2,
\end{equation}
whose solution, for a generic value of the topological mass
$\mu\ne\pm{l}^{-1}$, reads
\begin{subequations}\label{eq:minsolTMG}
\begin{eqnarray}
ds^2&=&\frac{l^2}{y^2}\left\{-\left[F_1(u)\left(\frac{y}l\right)^{1-l\mu}
+\frac{\kappa{l}\mu}{1+l\mu}\left(\frac{\mathrm{d}\Phi}{\mathrm{d}u}\right)^2
\ln{\left(\frac{y}l\right)}y^2\right]du^2-2dudv+dy^2\right\},\quad~\\
\Phi&=&\Phi(u).
\end{eqnarray}
\end{subequations}
It easy to note that the above configuration is just a superposition
of the vacuum AdS wave of topologically massive gravity
(\ref{eq:TMGAdSwv}) and the AdS wave of Einstein gravity obtained by
considering a minimally coupled scalar field as a source
(\ref{eq:minsol}) with an effective gravitational constant
$\kappa_{\mathrm{eff}}=\kappa/[1+(l\mu)^{-1}]$. The structural
function satisfies
\begin{equation}\label{eq:K_Gmueff}
\Box{F}={\mu_{\mathrm{eff}}}^2F+(\ldots),
\end{equation}
where the terms within $(\ldots)$ depend nonlinearly on $F$. Here
the effective topological mass is the same than in the vacuum case,
i.e. $\mu_{\mathrm{eff}}=\mu\sqrt{1-(l\mu)^{-2}}$.

For the critical values of the topological mass $\mu=l^{-1}$ and
$\mu=-l^{-1}$, we have the following solutions
\begin{subequations}\label{eq:minsolTMGl1}
\begin{eqnarray}
ds^2&=&\frac{l^2}{y^2}\left\{-\left[F_1(u)
+\frac\kappa2\left(\frac{\mathrm{d}\Phi}{\mathrm{d}u}\right)^2
y^2\right]\ln{\left(\frac{y}l\right)}du^2
-2dudv+dy^2\right\},\\
\Phi&=&\Phi(u),
\end{eqnarray}
\end{subequations}
and, respectively
\begin{subequations}\label{eq:minsolTMGl-1}
\begin{eqnarray}
ds^2&=&\frac{l^2}{y^2}\left\{-\left[F_1(u)-\frac{\kappa{l}^2}2
\left(\frac{\mathrm{d}\Phi}{\mathrm{d}u}\right)^2
\ln{\left(\frac{y}l\right)}\right]\ln{\left(\frac{y}l\right)}\frac{y^2}{l^2}du^2
-2dudv+dy^2\right\},\quad~\\
\Phi&=&\Phi(u).
\end{eqnarray}
\end{subequations}

\subsection{AdS waves for free nonminimally coupled scalar fields}

We turn our attention now to the free scalar fields allowing
nonminimal coupling ($\lambda=0$ and $\xi\ne0$) to an AdS wave. In
this case the pure radiation constraints imply that the mass of this
scalar field is fixed in terms of the nonminimal coupling by
Eq.~(\ref{eq:m^2}). In order to determine the AdS background allowed
by this source it is useful to consider the following redefinitions
\begin{subequations}\label{eq:xH2yFTMG}
\begin{eqnarray}
x&=&\frac{(1-4\xi)l\mu}4\left(\frac{y}{lf}\right)^{4\xi/(1-4\xi)}, \\
H(u,x)&=&
\frac{l^2f^2}{y^2}F+l^2f\frac{\mathrm{d}^2f}{\mathrm{d}u^2},
\end{eqnarray}
\end{subequations}
for which the $uu-$equation can be rewritten as
\begin{equation}\label{eq:ghypergeomDETMG}
x^2\partial^3_{xxx}H-\frac{x[4\xi{x}-(1-4\xi)l\mu-3]}{4\xi}\partial^2_{xx}H
-\frac{4\xi{x}-(1-2\xi)[(1-4\xi)l\mu+1]}{8\xi^2}\partial_xH
-\frac{(1-4\xi)}{2\xi}H=0.
\end{equation}
We recognize the generalized hypergeometric differential equation
whose general solution is expressed as
\begin{eqnarray}\label{eq:solHTMG}
H(u,x)&=&F_0(u)\left(\frac{4x}{(1-4\xi)l\mu}\right)^{-(1-4\xi)/(2\xi)}
        +F_1(u)\,{_2\tilde{F}_2}\!\left(1,\frac{1-4\xi}{2\xi};
        \frac{1-2\xi}{2\xi},\frac{1+(1-4\xi)l\mu}{4\xi};x\right)
                            \nonumber\\
      & &{}+F_3(u)\left(\frac{4x}{(1-4\xi)l\mu}\right)^
            {-(1-4\xi)(1+l\mu)/(4\xi)}
         {_1\tilde{F}_1}\!\left(\frac{(1-4\xi)(1-l\mu)}{4\xi};
                            \frac{1-(1-4\xi)l\mu}{4\xi};x\right),\nonumber\\
        \qquad~
\end{eqnarray}
where $F_0$, $F_1$, and $F_3$ are integration functions and
${_1\tilde{F}_1}(a;b;x)$ and ${_2\tilde{F}_2}(a,b;c,d;x)$ denote the
corresponding generalized hypergeometric functions
\cite{Erdelyi:1953}. Returning to the original variables
(\ref{eq:xH2yFTMG}) and using the coordinate transformation
(\ref{eq:formI}) it is possible to fix the function $f$ to the unity
and the function $F_0$ to zero. The resulting AdS wave configuration
for a free nonminimally coupled scalar field in topological massive
gravity is
\begin{subequations}\label{eq:solTMG}
\begin{eqnarray}
ds^2&=&\frac{l^2}{y^2}\Biggl\{
       -\biggl[F_1(u)\,{_2\tilde{F}_2}\!\left(1,\frac{1-4\xi}{2\xi};
                    \frac{1-2\xi}{2\xi},\frac{1+(1-4\xi)l\mu}{4\xi};
              \frac{(1-4\xi)l\mu\kappa\Phi^2}4\right)\frac{y^2}{l^2}
       \nonumber\\
    & &\qquad{}+F_3(u)\,{_1\tilde{F}_1}\!\left(\frac{(1-4\xi)(1-l\mu)}{4\xi};
                                           \frac{1-(1-4\xi)l\mu}{4\xi};
                                \frac{(1-4\xi)l\mu\kappa\Phi^2}4\right)
       \left(\frac{y}l\right)^{1-l\mu}\biggr]du^2
       \nonumber\\
    & &\qquad{}-2dudv+dy^2\Biggl\},\qquad~\\
\Phi&=&\frac1{\sqrt{\kappa}}
       \left(\frac{y}{l}\right)^{2\xi/(1-4\xi)},
\end{eqnarray}
\end{subequations}
where the functions $F_1$ and $F_3$ have been rescaled
appropriately. Here the case $\xi=1/4$ is excluded since, as in the
Einstein gravity case, it does not allow free configurations.

\subsection{Small topological mass limit: AdS waves for
Conformal Gravity}

Conformal gravity is a three-dimensional gravity theory just rigged
by the Cotton tensor. In
Refs.~\cite{Deser:2004wd,Ayon-Beato:2004fq,Ayon-Beato:2005bm,Jackiw:2005an}
the \emph{pp} waves of this theory with nonminimally coupled scalar
sources have been studied. We analyze the AdS wave solutions of this
theory using the fact that conformal gravity can be obtained from
topologically massive gravity in the limit of small topological
mass. More precisely, for $\mu\to 0$ with $\mu\kappa\sim 1$ the
topologically massive gravity equations reduce to
\begin{eqnarray}\label{eq:CG}
C_{\alpha\beta}=\tilde{\kappa}T_{\alpha\beta},
\end{eqnarray}
where $\tilde{\kappa}=\mu\kappa$ is a dimensionless gravitational
constant. The idea is to apply the above limit to the topologically
massive gravity AdS waves obtained previously in order to obtain the
corresponding AdS waves of conformal gravity.

In the minimal case $\xi=0$, the AdS wave configurations
(\ref{eq:minsolTMG}) at the small topological mass limit yields to
\begin{subequations}\label{eq:minsolCG}
\begin{eqnarray}
ds^2&=&\frac{l^2}{y^2}\left\{-\left[F_1(u)
+\tilde{\kappa}{l^2}\left(\frac{\mathrm{d}\Phi}{\mathrm{d}u}\right)^2
\ln{\left(\frac{y}l\right)}y\right]\frac{y}ldu^2-2dudv+dy^2\right\},\\
\Phi&=&\Phi(u).
\end{eqnarray}
\end{subequations}
On the other hand, one can solve the conformal gravity (\ref{eq:CG})
and conclude that the configuration (\ref{eq:minsolCG}) is the
general solution for a minimally coupled scalar field.

A similar construction is applied in the case of a free nonminimally
coupled scalar field ($\lambda=0$). At the small topological mass
limit one obtains from the solution (\ref{eq:solTMG}),
\begin{subequations}\label{eq:solCG}
\begin{eqnarray}
ds^2&=&\frac{l^2}{y^2}\Biggl\{
       -\biggl[F_1(u)\,{_2\tilde{F}_2}\!\left(1,\frac{1-4\xi}{2\xi};
                    \frac{1-2\xi}{2\xi},\frac{1}{4\xi};
              \frac{(1-4\xi)l\tilde{\kappa}\Phi^2}4\right)\frac{y}l
       \nonumber\\
    & &\qquad{}+F_3(u)\,{_1\tilde{F}_1}\!\left(\frac{1-4\xi}{4\xi};
                                           \frac{1}{4\xi};
                                \frac{(1-4\xi)l\tilde{\kappa}\Phi^2}4\right)
       \biggr]\frac{y}ldu^2
       -2dudv+dy^2\Biggl\},\qquad~\\
\Phi&=&\frac1{\sqrt{l\tilde{\kappa}}}
       \left(\frac{y}{l}\right)^{2\xi/(1-4\xi)},
\end{eqnarray}
\end{subequations}
which turns to be also the general solution of the conformal gravity
equations for a free nonminimally coupled scalar field.



\begin{thebibliography}{99}

\bibitem{Garcia:1981}
  A.~Garc\'{\i}a and J.~Pleba\'nski,
  J. Math. Phys. \textbf{22}, 2655 (1981).

\bibitem{Salazar:1983}
  H. Salazar, A.~Garc\'{\i}a, and J.~Pleba\'nski,
  J. Math. Phys. \textbf{24}, 2191 (1983).

\bibitem{Garcia:1983}
  A.~Garc\'{\i}a, Nuovo Cim. B \textbf{78}, 255 (1983).

\bibitem{Ozsvath:1985qn}
  I.~Ozsvath, I.~Robinson, and K.~Rozga,
  J.\ Math.\ Phys.\ \textbf{26}, 1755 (1985).

\bibitem{Kundt:1961}
  W.~Kundt, Z.~Phys. \textbf{163}, 77 (1961).

\bibitem{Robinson:1960}
  I.~Robinson and A.~Trautman,
  Phys. Rev. Lett. \textbf{4}, 431 (1960).

\bibitem{Bicak:1999h}
  J.~Bicak and J.~Podolsky,
  J.\ Math.\ Phys.\ \textbf{40}, 4495 (1999)
  [arXiv:gr-qc/9907048];
  4506 (1999)
  [arXiv:gr-qc/9907049].

\bibitem{Siklos:1985}
  S.T.C.~Siklos,
  in: \emph{Galaxies, axisymmetric systems and relativity},
  ed. M.A.H. MacCallum (Cambridge Univ. Press, Cambridge 1985).

\bibitem{Podolsky:1999sw}
  J.~Podolsky and J.~B.~Griffiths,
  Phys.\ Lett.\ A \textbf{261}, 1 (1999)
  [arXiv:gr-qc/9908008].

\bibitem{Podolsky:1997ik}
  J.~Podolsky,
  Class.\ Quant.\ Grav.\ \textbf{15}, 719 (1998)
  [arXiv:gr-qc/9801052].

\bibitem{Brown:1986nw}
  J.~D.~Brown and M.~Henneaux,
  Commun.\ Math.\ Phys.\ \textbf{104}, 207 (1986).

\bibitem{Banados:1999tw}
  M.~Ba\~nados, A.~Chamblin, and G.~W.~Gibbons,
  Phys.\ Rev.\ D \textbf{61}, 081901 (2000)
  [arXiv:hep-th/9911101].

\bibitem{Banados:wn}
  M.~Ba\~nados, C.~Teitelboim, and J.~Zanelli,
  Phys.\ Rev.\ Lett.\ \textbf{69}, 1849 (1992)
  [arXiv:hep-th/9204099].

\bibitem{Banados:1992gq}
  M.~Ba\~nados, M.~Henneaux, C.~Teitelboim, and J.~Zanelli,
  Phys.\ Rev.\ D \textbf{48}, 1506 (1993)
  [arXiv:gr-qc/9302012].

\bibitem{Deser:2004wd}
  S.~Deser, R.~Jackiw, and S.~Y.~Pi,
  Acta Phys.\ Polon.\ B \textbf{36}, 27 (2005)
  [arXiv:gr-qc/0409011].

\bibitem{Ayon-Beato:2004fq}
  E.~Ay\'{o}n--Beato and M.~Hassa\"{\i}ne,
  Annals Phys. \textbf{317}, 175 (2005)
  [arXiv:hep-th/0409150].

\bibitem{Ayon-Beato:2005bm}
  E.~Ay\'{o}n--Beato and M.~Hassa\"{\i}ne,
  Phys.\ Rev.\ D \textbf{71}, 084004 (2005)
  [arXiv:hep-th/0501040].

\bibitem{Jackiw:2005an}
  R.~Jackiw,
  ``Weyl invariant dynamics in 3 dimensions,''
  arXiv:gr-qc/0509035.

\bibitem{Stephani:2003tm}
  H.~Stephani, D.~Kramer, M.~MacCallum, C.~Hoenselaers, and E.~Herlt,
  \emph{Exact solutions of Einstein's field equations}
  (Cambridge University Press, Cambridge 2003).

\bibitem{Ayon-Beato:2004ig}
  E.~Ay\'{o}n--Beato, C.~Mart\'{\i}nez, and J.~Zanelli,
  ``Stealth Scalar Field Overflying a $2+1$ Black Hole,''
  arXiv:hep-th/0403228.

\bibitem{Ayon-Beato:2005tu}
  E.~Ay\'{o}n--Beato, C.~Mart\'{\i}nez, R.~Troncoso, and J.~Zanelli,
  Phys.\ Rev.\ D \textbf{71}, 104037 (2005)
  [arXiv:hep-th/0505086].

\bibitem{Ayon-Beato:2005c}
  E.~Ay\'{o}n--Beato, C.~Mart\'{\i}nez, R.~Troncoso, and J.~Zanelli,
  ``Stealth Scalar Field on a $D$--dimensional Generalized AdS space,''
  in preparation (2005).

\bibitem{Erdelyi:1953}
  A.~Erd\'elyi \emph{et al.},
  \emph{Higher Transcendental Functions} (McGraw-Hill, New York, 1953) Vol. 1.

\bibitem{Deser:1981wh}
  S.~Deser, R.~Jackiw, and S.~Templeton,
  Annals Phys. \textbf{140}, 372 (1982) [Erratum-ibid. \textbf{185}, 406 (1988)];
  Phys.\ Rev.\ Lett.\ \textbf{48}, 975 (1982).

\end{thebibliography}
\end{document}